\documentclass[10pt,preprint,twocolumn]{aastex7}
\newcommand{\roma}[1]{\uppercase\expandafter{\romannumeral#1}}
\newcommand{\speed}[1]{#1 km~s${}^{-1}$}
\newcommand{\accel}[1]{#1 km~s${}^{-2}$}

\newcommand{\nfig}[1]{Figure~\ref{#1}}

\usepackage{amsmath}
\usepackage{subfigure}
\usepackage{hyperref}

\usepackage{lineno}
\usepackage{color}
\usepackage{makecell}
\usepackage{graphicx}
\usepackage{natbib}
\usepackage{amssymb,txfonts}
\usepackage{multirow}
\usepackage{array}
\usepackage{sidecap}
\usepackage{hyperref}
\usepackage{epstopdf}
\usepackage{footnote}
\usepackage{tabularx}
\usepackage{booktabs}

\shorttitle{Formation of Double-decker Filament Through Component Magnetic Reconnection}
\shortauthors{Liu et al.}
\received{February 09, 2025}
\revised{April 30, 2025}
\accepted{May 19, 2025}

\begin{document}
\title{Deciphering the Formation and Dynamics of Double-decker Filament Through Component Magnetic Reconnection}
\correspondingauthor{Yuandeng Shen \& Yi Bi}

\author{Dongxu Liu$^{1,4}$}
\noaffiliation{}
\affiliation{Yunnan Observatories, Chinese Academy of Sciences, Kunming 650216, China}
\affiliation{State Key Laboratory of Solar Activity and Space Weather, School of Aerospace, Harbin Institute of Technology, Shenzhen 518055, China}
\affiliation{Shenzhen Key Laboratory of Numerical Prediction for Space Storm, Harbin Institute of Technology, Shenzhen 518055, China}
\affiliation{University of Chinese Academy of Sciences, Beijing, 100049, China}
\email{liudongxu@ynao.ac.cn}  

\author[orcid=0000-0001-9493-4418]{Yuandeng Shen}
\affiliation{State Key Laboratory of Solar Activity and Space Weather, School of Aerospace, Harbin Institute of Technology, Shenzhen 518055, China}
\affiliation{Shenzhen Key Laboratory of Numerical Prediction for Space Storm, Harbin Institute of Technology, Shenzhen 518055, China}
\email[show]{ydshen@hit.edu.cn}

\author[orcid=0000-0002-5302-3404]{Yi Bi}
\affiliation{Yunnan Observatories, Chinese Academy of Sciences, Kunming 650216, China}
\affiliation{University of Chinese Academy of Sciences, Beijing, 100049, China}
\email[show]{biyi@ynao.ac.cn}

\author{Zehao Tang}
\affiliation{Yunnan Observatories, Chinese Academy of Sciences, Kunming 650216, China}
\affiliation{University of Chinese Academy of Sciences, Beijing, 100049, China}
\email{tangzh@ynao.ac.cn}

\author{Chengrui Zhou}
\affiliation{Yunnan Observatories, Chinese Academy of Sciences, Kunming 650216, China}
\affiliation{University of Chinese Academy of Sciences, Beijing, 100049, China}
\email{zhouchengrui@ynao.ac.cn}

\author{Surui Yao}
\affiliation{Yunnan Observatories, Chinese Academy of Sciences, Kunming 650216, China}
\affiliation{University of Chinese Academy of Sciences, Beijing, 100049, China}
\email{yaosurui@ynao.ac.cn}

\begin{abstract}
The formation of double-decker filaments has long been an enigma in the field of solar physics. Using stereoscopic observations from the Solar Dynamics Observatory and the Solar Terrestrial Relations Observatory, we show that the double-decker filament formed on 2013 August 30 resulted from the splitting of a braided magnetic flux rope. The splitting was driven by component magnetic reconnection between intertwined field lines, triggered by the rotational motion in a part of one filament footpoint. This mechanism, inferred from observed small jets, brightenings, and bidirectional mass flows, differs from the previous conclusion attributing filament splitting to magnetic reconnection between the legs of confining magnetic field lines within or above the filament. The splitting speed might be modulated by the reconnection speed, as evidenced by the correspondence between the filament's slow and fast rising phases and the intermittent and violent brightening stages. Following the splitting, the upper branch of the double-decker filament erupted as a coronal mass ejection (CME), giving rise to a {\em GOES} soft X-ray M1.2 flare. In conclusion, our observations present a new formation mechanism for double-decker filaments, and the subsequent partial eruption is likely attributable to the torus instability of the background coronal magnetic field. Moreover, the detection of small jets within the filament provides new insights into the role of component magnetic reconnection in localized coronal heating processes.
\end{abstract}
\keywords{Sun: activity --- Sun: flares --- Sun: filaments --- Sun: jets --- Sun: magnetic fields}

\section{Introduction}\label{intro}
Solar filaments are plasma containing magnetic flux ropes (MFR) arcing up from the solar surface and suspended in the solar atmosphere \citep{2014LRSP...11....1P}. Filaments appear as elongated dark features on the solar disk because they are cooler compared to the surrounding coronal material \citep[][]{2007ApJ...667L.105J}. Still, when they appear on the disk limb against the dark sky, they appear as giant bright loops and are called prominences \citep{2015ApJ...814L..17S, 2024ApJ...970..110G}. Therefore, filaments and prominences are the same entity but are observed from different angles. The eruption of solar filaments has attracted much attention in the past decades for the potential effects to cause coronal mass ejections (CMEs) and geomagnetic storms to disturb the near-Earth space environment \citep{2011LRSP....8....1C}. Previous studies suggest that filament eruptions could be divided into failed, partial, and successful eruptions, in which the partial and full types are associated with CMEs while the failed type is not. Although all types of filament eruptions are associated with flares, small flares with less energy seem to be more often accompanied by failed filament eruptions because their energy is insufficient to cause a CME \citep[][]{2011RAA....11..594S}. 

\cite{2001ApJ...549.1221G} summarized that filament eruptions are driven by the magnetic reconnection between the two legs of the confining magnetic field, where the location of the reconnection site determines the eruption types. Namely, when the reconnection site is located below, inside, and above a filament, the outcome of the eruption corresponds to a complete, partial, and failed filament eruption, respectively. It should be pointed out that the failed filament eruption discussed here only means the failure of the filament instead of the entire MFR system. For example, \cite{2012ApJ...750...12S} reported a failed filament eruption associated with a CME, where the reconnection location is evidenced above the filament but within the MFR. If the reconnection site is located within the filament, the latter will be separated into two branches to form a so-called double-decker filament \citep[e.g.,][]{2012ApJ...756...59L, 2018NewA...65....7T}. Therefore, the terms complete, partial, and failed eruptions are relative to the eruption of the filament rather than the whole MFR, and all eruption types are theoretically associated with CMEs. This is different from the concept mentioned in the previous paragraph. \cite{2006ApJ...637L..65G} modeled the partial eruption of an MFR. They found the three-dimensionality and the existence of dipped field lines grazing the photosphere \citep[i.e., a bald patch, which refers to a region on the polarity inversion line (PIL) where the photospheric magnetic field is tangential, the z-component of the magnetic induction is zero, and the magnetic field lines are concave upwards,][]{1993A&A...276..564T} are key to influencing the partial eruption of an MFR. The bald patch keeps part of the rope behind, while the three-dimensionality induces internal and external reconnections to cause the eruption of the other parts. \cite{2018ApJ...856...48C} reported partial filament eruptions due to the vertical splitting of an original filament into two parts by internal reconnection, and the authors claimed that their observations are consistent with the model proposed in \cite{2006ApJ...637L..65G}. According to this model, the internal reconnection-induced vertical splitting of an MFR should be accompanied by the occurrence of near-simultaneous flares (say, within a few minutes), and this has been evidenced in several recent studies \citep[e.g.,][]{2012ApJ...750...12S, 2018ApJ...856...48C, 2023ApJ...953..148S}.

In some particular events, a filament might be composed of different groups of threads whose degrees of twisting are very different. Due to some external or internal disturbances, the highly twisted filament threads might erupt because of kink instability, leaving behind the other less twisted filament threads remaining stable. This kind of filament eruption is sometimes also called a partial filament eruption \citep{2015ApJ...805...48B}. Partial filament eruptions can also originate from double-decker filament systems that consist of two MFRs separated in height or one MFR suspended above another sheared arcade structure along the same PIL \citep{2012ApJ...756...59L}. The stability of the two types of double-decker filament systems was verified through numerical simulations \citep{2014ApJ...792..107K}. However, the formation mechanism of such a particular magnetic structure remains an open question. To date, several candidate formation mechanisms have been proposed in the literature. For example, the formation of double-decker filaments could be due to the rising of chromospheric fibrils from below a pre-existing one \citep{2014ApJ...789..133Z, 2020ApJL...898...L12}, or by successive confined tether-cutting like eruptions at the same PIL \citep{2024ApJ...964..125S}, or by splitting a single filament into two branches through magnetic reconnection within the filaments between the two legs of the confining magnetic field lines \citep[e.g.,][]{2012ApJ...750...12S, 2018ApJ...856...48C, 2023ApJ...953..148S}. In these possible mechanisms, the formed double-decker filament system is usually unstable and often erupts (partially) within a short timescale during or after the formation. Therefore, some authors called such kind of flux rope system separated in height as transient double-decker filaments \citep{2024ApJ...964..125S}. The double-decker filaments mentioned above typically have the same sign of helicity because they are formed successively through the same physical process within the same magnetic region. However, some observations indicated that some double-decker filaments show a novel magnetic structure whose two branches manifest opposite magnetic helicities \citep[e.g.,][]{2019ApJ...872..109A, 2021ApJ...923..142C, 2023ApJ...959...69H}. \cite{2023ApJ...959...69H} proposed that such a particular magnetic structure of double-decker filaments might be composed of two branches originating from different magnetic regions, and this type of double-decker filaments might be more stable than those with the same helicity sign \citep{2021ApJ...909...32P}. Many observational studies have investigated the eruptive behaviors of double-decker filaments in recent years. \cite{2012ApJ...756...59L} found that the rising of the lower filament interacts with the upper filament, transferring flux and current can make the unstable and even eruption of the upper filament when its axial flux reaches a critical value \citep{2020ApJL...898...L12}. This scenario primarily involves magnetic reconnection between the two filaments consisting of the double-decker filament, and it has been supported by a few further observations \citep{2014SoPh..289..279Z, 2024Univ...10...42L}.

\begin{figure*}[thbp]
\centering
\includegraphics[width=0.8\textwidth]{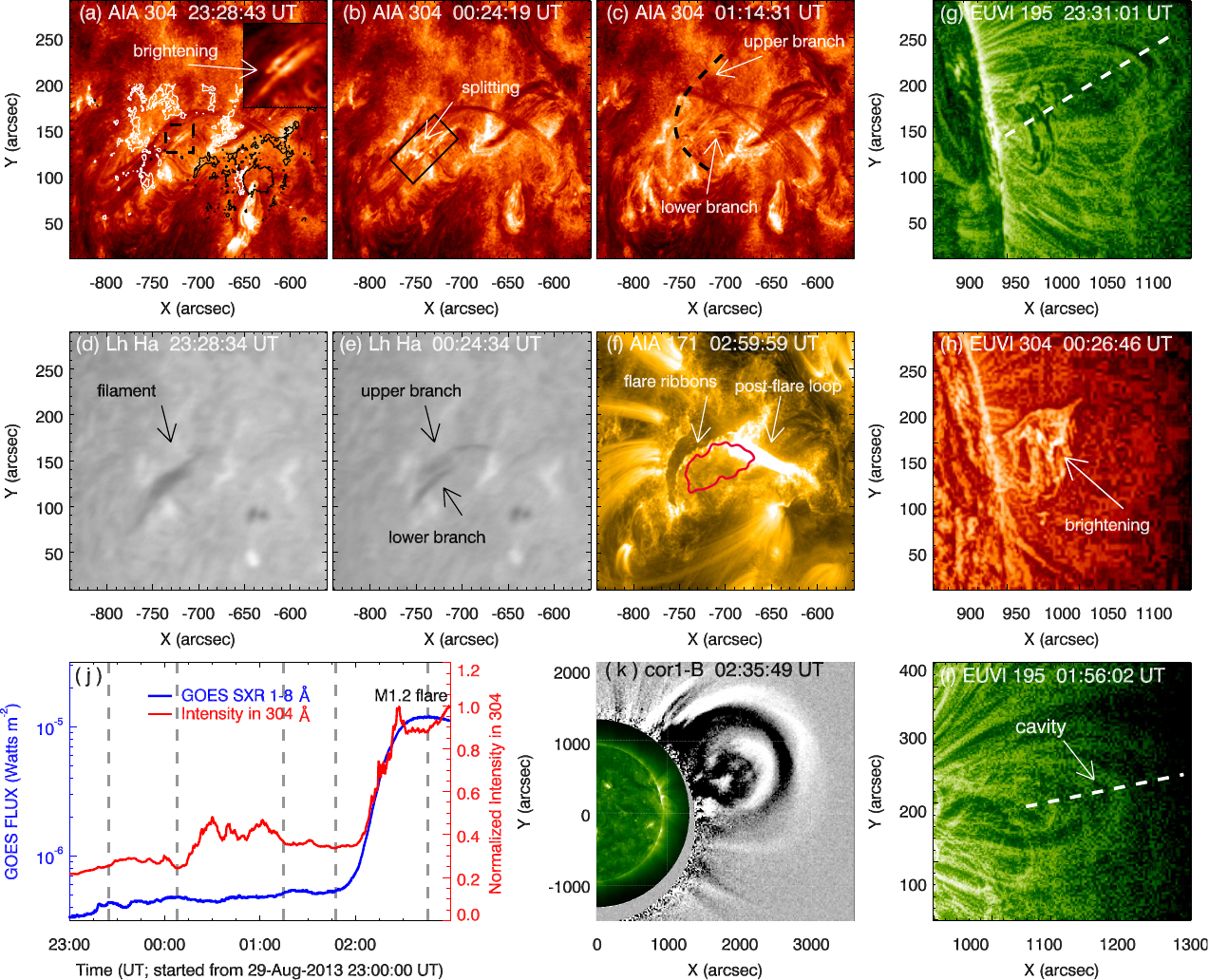}
\caption{An overview of the partial eruption is illustrated in this figure. Panel (a) shows an overview of the active region in 304 \AA, with contours in black and white representing the negative and positive polarities of the magnetic field, respectively. Panel (b) captures the splitting of the filament, highlighted by the black box, which marks the region selected for calculating the light curve in panel (j). Panel (c) depicts the rise of the upper branch of the filament post-splitting. The dashed line represents the path of the time-distance plot, as shown in \nfig{fig2}(a). Panels (d) -- (e) display the splitting of the filament in the H$\alpha$ images acquired by the Learmonth Solar Observatory. Panel (f) features the flare ribbon and a post-flare loop in the 171 \AA\ image. The red contours represent the 3000-count isolevel in the 6–12 keV range of the flare image from RHESSI. Panels (g) -- (i) display the splitting of the filament and the rise of the cavity in the {\em STEREO}-B EUVI 304 \AA\ and 195 \AA\ images. In panels (g) and (i), the dashed lines indicate the trajectory used for the time-distance plot, as depicted in \nfig{fig2}(b) and (c). Panel (j) presents the {\em GOES} SXR 1–8 \AA\ curve and the AIA 304 \AA\ light curve from 00:00 to 04:00 UT. Panel (k) displays differential images of a halo CME detected by {\em STEREO} COR1-B. An animation of the {\em SDO}/AIA 304 \AA\ is available. Its duration is 74 s and it covers the period from 22:30 UT on August 29 to 03:30 UT on August 30.} 
\label{fig1}
\end{figure*}

Using multi-wavelength observations taken by the {\em Solar Dynamics Observatory} \citep[{\em SDO};][]{2012SoPh..275....3P} and the {\em Solar Terrestrial Relations Observatory} \citep[{\em STEREO};][]{2008SSRv..136....5K}, we present observations of a new type of filament splitting and the subsequent formation of a transient double-decker filament. Our analysis results provide new clues for understanding the formation of double-decker filaments and the associated partial filament eruptions. In this paper, we focus on the 304 \AA\ and 171 \AA\ images from the Atmospheric Imaging Assembly \citep[AIA;][]{2012SoPh..275...17L} onboard the {\em SDO}, which have a pixel size of $0\arcsec.6$ and a cadence of 12 seconds. The line-of-sight (LOS) magnetograms from the Helioseismic and Magnetic Imager \citep[HMI;][]{2014SoPh..289.3483H} onboard {\em SDO} has a pixel size of $0\arcsec.6$ and a cadence of 45 seconds. We also used the 304 \AA\ and 195 \AA\ images taken by the Extreme Ultraviolet Imager \citep[EUVI;][]{2004SPIE.5171..111W} onboard the {\em STEREO}-B. Both channels have a pixel size of $1\arcsec.59$, but their cadences are of 10 and 5 minutes, respectively. Additionally, the COR1 instrument onboard the {\em STEREO}-B captures white-light images of the inner solar corona, covering a range from 1.4 to 4 $R_{\odot}$ \citep{2003SPIE.4853....1T}. The COR1 images are captured every 5 minutes and feature a pixel resolution of $15\arcsec$. We also utilize the H$\alpha$ images obtained from the Learmonth Solar Observatory, which is part of the Global Oscillation Network Group. The H$\alpha$ data have a cadence of 1 min and a pixel size of $1\arcsec.05$ \citep{1994SoPh..152..321H,1994SoPh..152..351H}.

\section{Results}\label{sec:result}
The filament, on 2013 August 29, was located in NOAA active region AR11836 on the northeast district of the solar disk from the Earth's view angle. As displayed in the top row of \nfig{fig1}, an S-shaped filament can be distinctly observed in the AIA 304 \AA\ images. In \nfig{fig1}(a), the white and black contours denote the $\pm 100$ Gauss boundaries of the positive and negative polarities as observed in the HMI LOS magnetogram. It suggests that the filament was situated along the PIL between the opposite magnetic polarities, and the west and east footpoints were rooted in negative and positive polarities, respectively. At about 23:28:43 UT on August 29, 2013, a brightening appeared in the filament at the heliospheric coordinate (in arcseconds) of about (x, y) = [-720, 140]. The brightening first happened intermittently from 23:28:43 UT on August 29 to 00:08:31 UT on August 30 and grew in size and brightness. Then, the brightening underwent a rapid increase in area and brightness around the southeast section of the filament, which reached the maximum brightness at about 00:24:19 UT and disappeared at about 01:10:55 UT on August 30. The lightcurve within the brightening region (see the black box in \nfig{fig1}(b)) is plotted in \nfig{fig1}(j) as a red curve, from which one can see the small peaks caused by the intermittent brightenings from 23:28:43 UT on August 29 to 00:08:31 UT on August 30 and the strong peak caused by the violent brightening from 00:08:31 UT to 01:10:55 UT on August 30. The brightening lasted for more than one and a half hours, during which the filament evolved into a double-decker filament consisting of two branches (see the arrows in \nfig{fig1}(c) and (e), and the animation available in the online journal). After the formation of the double-decker filament, the upper branch rose slowly and rotated clockwise when looking along the filament axis from the west footpoint, and it finally violently erupted and caused a {\em GOES} M1.2 flare as shown by the soft X-ray flux at the energy bands of 1 -- 8 \AA\ in \nfig{fig1}(j). The flare ribbon and loops at 02:59:59 UT in association with the partial eruption of the newly formed double-decker filament are displayed in \nfig{fig1}(f), in which the red contour represents the {\em RHESSI} X-ray source at the energy band of 6–12 keV. In contrast, the lower branch remained stable and gradually became invisible in the AIA images (see the animation available in the online journal).

\begin{figure*}[thbp]
\centering
\includegraphics[width=0.8\textwidth]{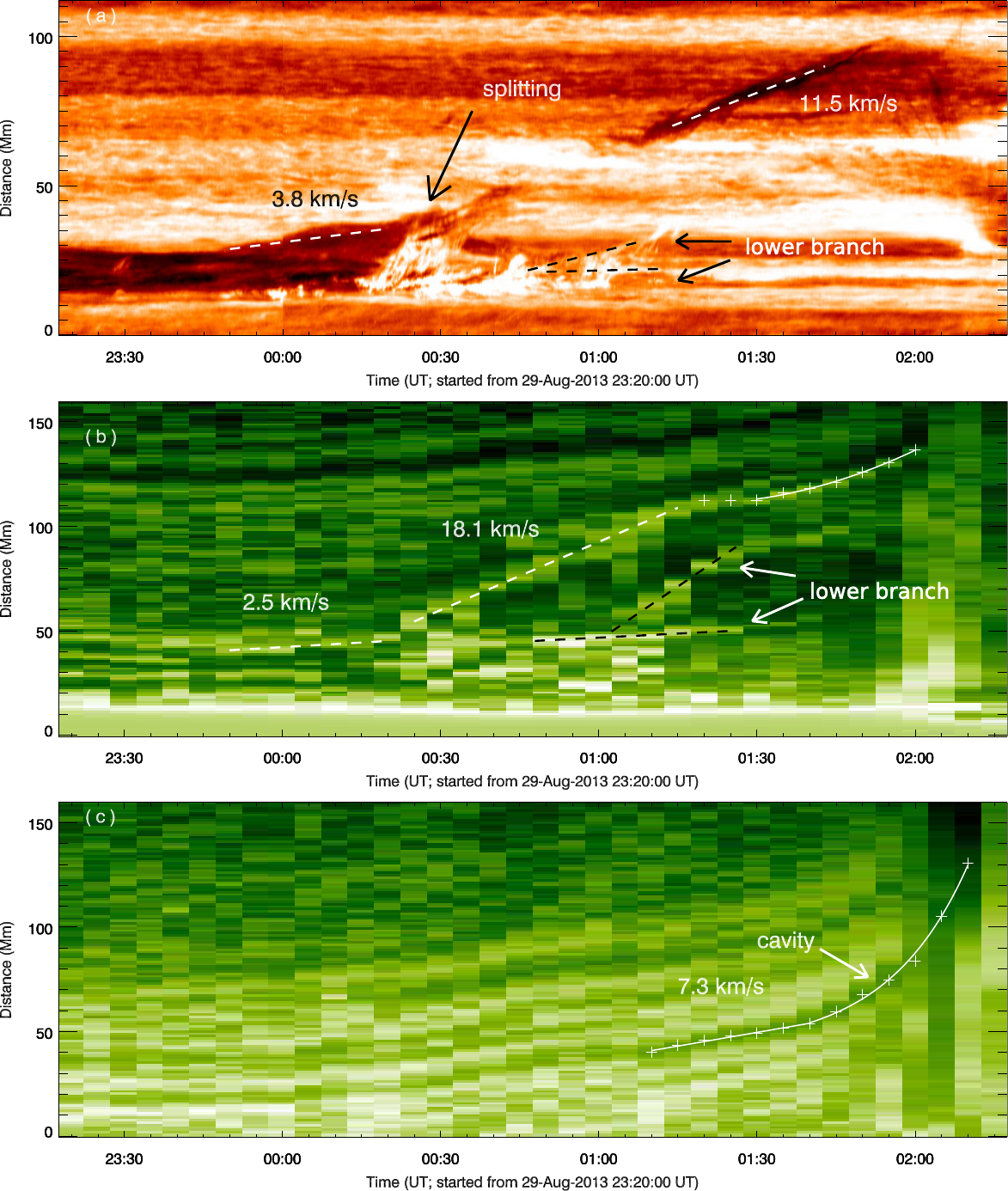}
\caption{
Panels (a) and (b) display the time–distance diagrams derived from the AIA 304 \AA\ and {\em STEREO}-B 195 \AA\ images. The dashed lines trace the movement of the upper branch. Panel (c) displays the time–distance plot, which was used to track the kinematic evolution of the erupting cavity.
}
\label{fig2}
\end{figure*}

\begin{figure*}[thbp]
\centering
\includegraphics[width=0.85\textwidth]{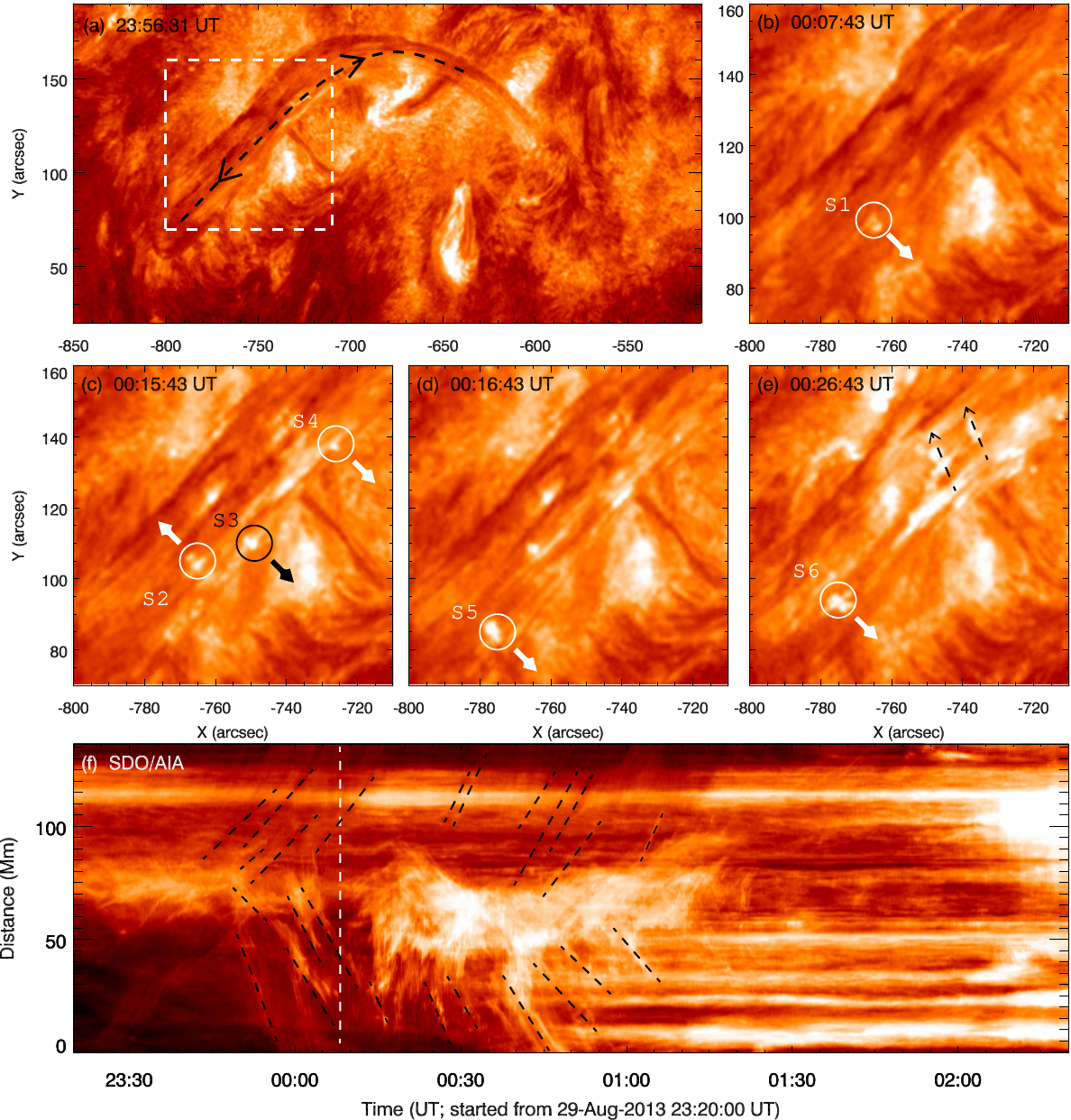}
\caption{The AIA 304 \AA\ images illustrate the splitting of the filament and several small jets observed during the splitting process. The white box in panel (a) represents the field of view (FOV) shown in panels (b) -- (e), while the dashed line indicates the path used for the time-distance plot in panel (f). Panels (b) -- (e) highlight some small jets observed during the splitting process, labeled as S1 -- S6. In panel (e), the dashed lines indicate the movement of brightened plasma blobs perpendicular to the filament axis. In panel (f), the black dashed lines show the bidirectional flow of plasma along the filament towards both footpoints, while the white dashed line indicates the structure as observed at 00:08 UT. 
}
\label{fig3}
\end{figure*}

The {\em STEREO}-B/EUVI 304 \AA\ and 195 \AA\ images are displayed in \nfig{fig1}(g) -- (i). At 00:30 UT on August 30, 2013, the separation angle between {\em STEREO}-B and Earth was $138.056\degr$. Therefore, the eruption source region was slightly behind the west limb of the solar disk when observing from the {\em STEREO}-B, and the filament can be observed as a prominence above the disk limb (see \nfig{fig1}(g)). Here, the low-resolution EUVI images are further processed using the multi-scale Gaussian normalization (MGN; \citep{2014SoPh..289.2945M}) technique to better display the prominence. At 00:26:46 UT, the brightening feature observed in the AIA 304 \AA\ images can also be observed in the prominence in the {\em STEREO}-B/EUVI 304 \AA\ image (see \nfig{fig1}(h)). About 43 minutes after the end of the brightening in the filament, the {\em GOES} X-ray M1.2 flare started at approximately 01:54 UT and peaked at 02:46 UT (\nfig{fig1}(j)). Correspondingly, {\em STEREO} 195 \AA\ observations show the rise of the prominence's upper branch and the overlying coronal cavity. By 01:55 UT, the prominence becomes nearly indiscernible, likely due to heating and subsequent mass drainage, while the cavity exhibits significant acceleration (see \nfig{fig1}(i)). The eruption of the upper branch of the newly formed double-decker filament also triggered a halo CME, which propagated northwestward at a speed of \speed{830} as observed in COR1 images. This event was registered in the {\em SOHO}/LASCO halo CME catalog with a reported speed of \speed{949} and was further captured by the {\em STEREO}-B/COR1, exhibiting a typical three-part structure during its northwestward ejection (as shown in \nfig{fig1}(k)).

\nfig{fig2} (a) and (b) present the detailed evolution process of the filament, using time-distance plots made along paths parallel to the moving direction with the AIA 304 \AA\ and the {\em STEREO}-B 195 \AA\ images (see the dashed line in \nfig{fig1}(c) and (g)). Before the start of the brightening in the filament at about 23:28:43 UT on August 29, the filament kept stable, and no obvious movement could be detected. As the brightening activities occurred, the filament showed a gradual rise motion. Corresponding to the intermittent and violent brightening phases, the filament exhibited gradual inflation followed by the rise of its upper branch. During the intermittent brightening phase, the entire filament exhibited slow inflation with speeds of \speed{3.8} (AIA 304 \AA\ ) and \speed{2.5} (EUVI 195 \AA\ ). This was followed by a violent brightening phase, and during this phase, the upper branch rose at speeds of \speed{11.5} and \speed{18.1} in the respective wavelengths. The majority of the lower branch maintained its original height while exhibiting persistent brightening. Around 01:00 UT, a substructure with discernible rising motion split from it, revealing the dynamical complexity of the lower branch. The brightenings can clearly be identified in the AIA 304 \AA\ time-distance plot, and those during the intermittent and violent brightening phases are indicated by the white dashed lines and black arrow in \nfig{fig2}(a), respectively. The brightenings during the intermittent phase can not be observed in the {\em STEREO}-B 195 \AA\ time-distance plot due to the low spatiotemporal resolution of the images. After the two phases of the brightenings, the filament was split into two branches as a double-decker filament. The upper branch of the double-decker filament remained in a low-speed state for about 15 minutes starting at approximately 01:15 UT (see the thick plus signs in \nfig{fig2}(b)) before it began to accelerate upward with an slow acceleration of about \accel{0.008} (see the white parabola in \nfig{fig2}(b)). This is consistent with the rise and rotation of the filament observed in the AIA 304 \AA\ images. Meanwhile, as the filament gradually disappeared in both AIA and {\em STEREO}-B observations, we monitored the overlying cavity to track the subsequent eruption. The time-distance plot of the cavity along the dashed line marked in \nfig{fig1}(i) is presented in \nfig{fig2}(c). Before 01:40 UT, the cavity rose by only 1–3 pixels every 5 minutes, exhibiting a linear trend with a fitted velocity of \speed{7.3}. After 01:40 UT, however, the cavity underwent a violent acceleration, which was best fitted by an exponential function (see the white curve in \nfig{fig2}(c)).

The brightenings in the filament were associated with small jets perpendicular to the filament threads, and each small brightening was actually the eruption source location of a small jet. In the meantime, bidirectional mass flows originating from these brightenings and moving along the filament threads were also observed in the AIA 304 \AA\ images (see the animation available in the online journal). \nfig{fig3} shows several small jets in enlarged images and the bidirectional mass flows in the time-distance plot made along the filament spine using the AIA 304 \AA\ images. \nfig{fig3}(a) shows the AIA 304 \AA\ image at 23:56:31 UT on August 29 when the bidirectional mass flows were prominent along the filament spine (see the arrows), in which the entire filament structure can well be observed and the overlaid dashed box indicates the field of view of the following panels from \nfig{fig3}(b) to (e). Although the weak brightenings during the intermittent brightening phase were all associated with small jets, we selected to show the stronger small jets during the violent brightening phase in detail using the AIA 304 \AA\ images. \nfig{fig3}(b) shows the first small brightening that represents the start of the violent brightening phase at 00:07:43 UT, in which the white arrow indicates the ejecting direction of the small jet. Then, multiple brightening points appeared on the filament threads, and all of them were associated with small jets. It is found that most of the small jets were southwest orientated, while a small fraction of them were ejected in the northeast direction (see the white circles and arrows in \nfig{fig3}(c) -- (e) and the animation available in the online journal). In addition, some heated plasma blobs were moved in the northeast direction perpendicular to the filament threads (see the black arrows in \nfig{fig3}(e)). However, these blobs are difficult to identify as discrete jet-like structures in the images, and therefore, no further analysis was conducted. A time-distance plot along the filament is created using the AIA 304 \AA\ images to study the kinematics of the bidirectional moving plasma flows (see \nfig{fig3}(f)). It is clear that the bidirectional mass flows originated from the brightening regions and can be divided into two phases, as indicated by the black dashed lines. The first (second) phase was associated with the intermittent (violent) brightening phase, which had an average speed of about \speed{57 (69)}. The faster speed of the bidirectional plasma flows might imply a stronger energy release process during the violent brightening phase.

\begin{figure*}[thbp]
\centering
\includegraphics[width=0.8\textwidth]{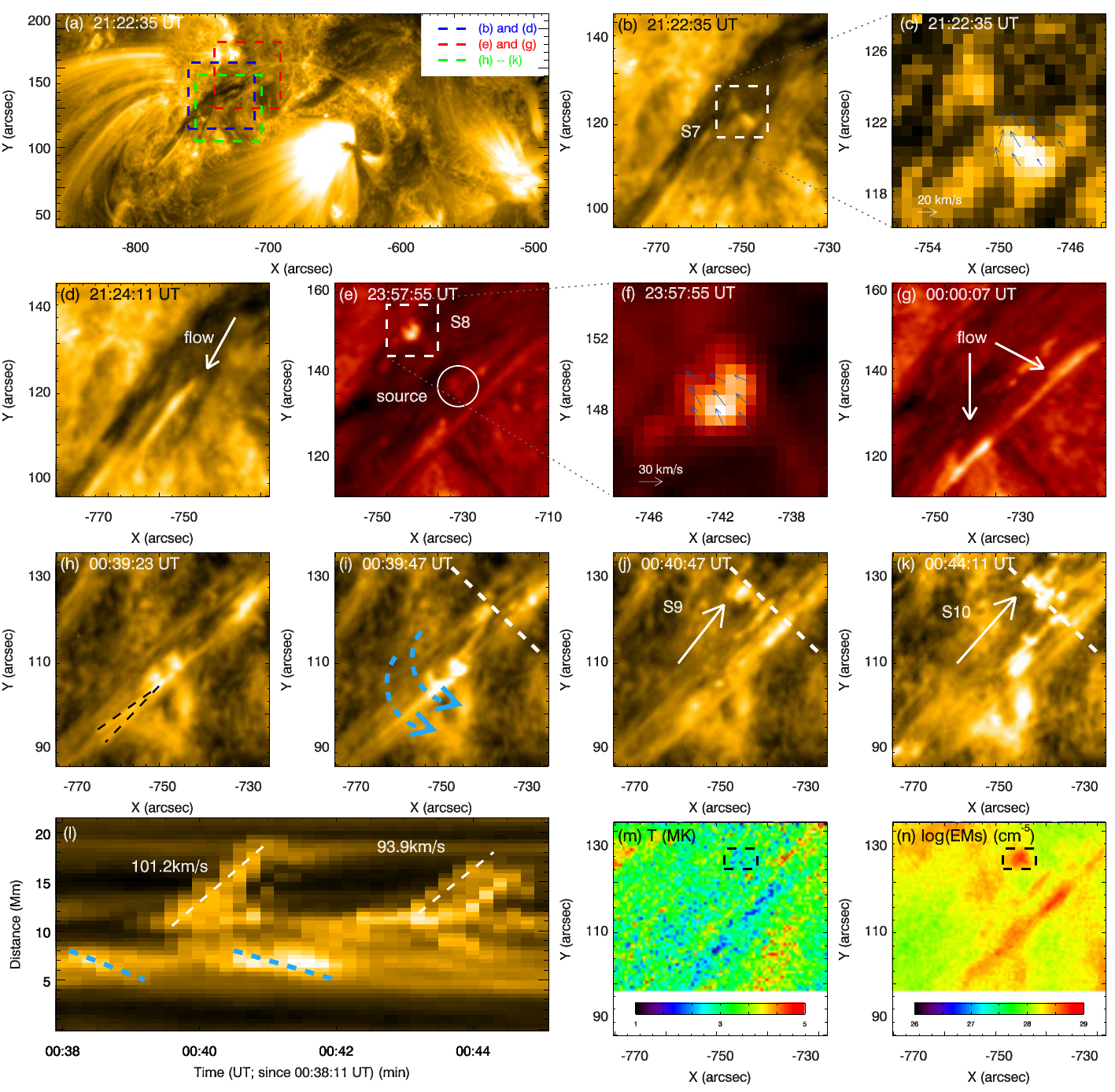}
\caption{The observations of small jets S7 -- S10 during the filament splitting process. Panel (a) displays an AIA 171 \AA\ image to show the FOV of the small jets. Panels (b) and (e) highlight the locations of small jets S7 and S8, as captured in AIA 171 \AA\ and 304 \AA\ images. The velocity fields of S7 and S8 are shown in panels (c) and (f), where the blue arrows represent the flow magnitudes and directions derived using the FLCT method. Panels (d) and (g) depict the brightened flows along the filament threads, which are associated with these small jets. Panel (h) displays the brightening of different clusters of magnetic field lines, indicated by dashed lines, which form a small angle, as observed at 00:39 UT. The component reconnection of these clusters led to S9 at 00:40 UT (see panels (i) -- (j)). The blue dashed lines in panel (i) illustrate the spinning motion of the brightened flow. Panel (k) shows S10 observed at 00:44 UT. Panel (l) presents the time-distance diagram along the dashed lines in panels (i) –- (k), which represent the trajectory of the small jets. The velocities of both small jets are shown in panel (l). Additionally, the blue dashed lines in panel (l) highlight the rotational motion of the brightened flow. Panels (m) and (n) display the weighted temperature and weighted emission measures of the region, with the scales shown below the figures. The black boxes in panels (k) and (i) indicate the presence of the S9 observed at 00:40 UT.}
\label{fig4}
\end{figure*}

\begin{deluxetable*}{cccccccccccc}
\tabletypesize{\scriptsize}
\tablecaption{Physical Properties of Observed Events \label{tab:table1}}
\tablehead{
\colhead{Number} & \colhead{Time} & \colhead{Length} & \colhead{Width} & \colhead{Density} & \colhead{$T$} & \colhead{Velocity} & \colhead{L} & \colhead{$E_t$} & \colhead{$E_k$} & \colhead{$U$} & \colhead{$B_T$}\\ 
& \colhead{($\rm UT$)} & \colhead{($\rm Mm$)} & \colhead{($\rm Mm$)} & \colhead{($10^9~ \rm cm^{-3}$)} & \colhead{($\rm MK$)} & \colhead{($\rm km~s^{-1}$)} & \colhead{($\rm s$)} & \colhead{($10^{26}~ \rm erg$)} & \colhead{($10^{24}~ \rm erg$)} & \colhead{($\rm erg~cm^{-3}$)} & \colhead{($\rm Gauss$)} 
}
\colnumbers
\startdata
S1& Aug 30 07:43 & 4.3 & 3.4 & 7.0 & 2.84 & 92.3 & 48 & 
3.21 & 19.39 & 8.72 & 14.80 \\
S2 & Aug 30 15:43 & 3.7 & 2.8 & 8.0 & 3.00 & 57.7 & 96 & 2.25 & 5.05 & 10.12 & 15.94 \\
S3 & Aug 30 15:43 & 2.2 & 1.8 & 9.6 & 2.52 & 58.6 & 36 & 0.56 & 1.54 & 10.27 & 16.06 \\
S4 & Aug 30 15:43 & 1.9 & 1.6 & 8.8 & 3.91 & 31.2 & 48 & 0.55 & 0.28 & 14.39 & 19.01 \\
S5 & Aug 30 16:43 & 4.5 & 2.2 & 8.2 & 2.31 & 49.6 & 72 & 1.34 & 2.88 & 8.00 & 14.17 \\
S6 & Aug 30 26:43 & 3.4 & 2.2 & 8.5 & 2.40 & 113.1 & 36 & 1.09 & 11.72 & 9.32 & 15.30 \\
S7 & Aug 29 21:22 & 9.3 & 2.8 & 7.1 & 4.02 & 15.8 & - & 6.77 & 0.85 & 11.84 & 17.24 \\
S8 & Aug 29 23:57 & 16.1 & 3.7 & 7.0 & 2.37 & 30.7 & - & 11.89 & 9.52 & 6.93 & 13.19 \\
S9 & Aug 30 00:40 & 8.5 & 5.0 & 7.2 & 2.90 & 101.2 & 96 & 13.58 & 96.44 & 9.26 & 15.25\\
S10 & Aug 30 00:43 & 6.4 & 3.1 & 8.8 & 2.84 & 93.9 & 192 & 5.00 & 31.40 & 11.00 & 16.62 \\
\enddata
\tablecomments{
This table presents the measured and derived parameters of small jets. 
Here, $T$ represents the temperature, $E_t$ denotes the total thermal energy, 
$E_k$ refers to the total kinetic energy, $U$ stands for the energy density, 
and $B_T$ indicates the transverse magnetic field component involved in the reconnection process. S7 and S8 originated inside the filament and appeared as weak, discontinuous bright plasmoids. This made them difficult to track, and their lifetimes could not be determined.
}
\end{deluxetable*}

In addition to the studied small jets, we also observed some similar but relatively larger jet-like features. \nfig{fig4} shows a detailed analysis of these small jets, and they are named S7 -- S10 for convenience. \nfig{fig4}(a) displays an AIA 171 \AA\ image with a large field of view, in which the colored boxes indicate the field of views of the following panels from \nfig{fig4}(b) to (k). S7 and S8 occurred at about 21:22:35 UT and 23:57:55 UT on August 29, respectively. They ejected perpendicular to the filament threads in the northeast direction and were composed of discrete bright pixels (see \nfig{fig4}(b) and (e)). Notably, S8 transformed into brighter plasmoids after traversing some filament threads, which suggests that these jets originated inside the filament, with their extreme ultraviolet (EUV) emission appearing as weak and discontinuous bright plasmoids due to absorption by the foreground filament threads. We calculate the velocity field in the plane of the sky of S7 and S8 by using the Fourier Local Correlation Tracking (FLCT; \citep{2008ASPC..383..373F}) method with two sequential AIA images separated by 12 seconds. The velocity maps of S7 (S8) are shown in \nfig{fig4}(c) ((f)), which indicates that the velocity of S7 (S8) was about \speed{15.8 (30.7)}. We also calculated the velocities of the small jets studied in \nfig{fig3}, and their velocities generally ranged from \speed{30 to 110}. The low speeds of S7 and S8 are probably due to the higher altitude of the filament threads, which obscures the jets, making continuous tracking more difficult and, therefore, resulting in a larger error in the velocity measurement. In addition, the ejection direction relative to the observer could also be a possible reason. For S7 and S8, we also observed the subsequent corresponding brightening of filament threads at 21:24:11 UT on August 29 and 00:00:07 UT on August 30, respectively. The brightness of these filament threads is mainly due to the heated plasma moving along them, resembling the characteristics of post-reconnection loops generated by magnetic reconnections (see \nfig{fig4}(d) and (g)).

It is noted that S9 ejected perpendicular to the filament threads in the northeast direction and originated at the crossing of two misaligned filament threads, as indicated by the two black dashed lines in \nfig{fig4}(h). It occurred at about 00:40:47 UT on August 30, and the bright jet body across several filament threads manifested as discrete plasmoids (see \nfig{fig4}(j)). During the ejection process, rapid spinning of some filament threads was observed, and the direction of the rotation motion is indicated by the blue curved arrows in \nfig{fig4}(i). Several minutes after S9, we observed another similar small jet (S10) at about 00:44:11 UT in the same direction as S9 (see \nfig{fig4}(k)), which exhibited a brighter jet body than S9 but with a similar structure. A time-distance plot along the direction of the jets was made using the AIA 171 \AA\ images (\nfig{fig4}(l)). Based on the time-distance plot, the projected speed of S9 is measured to be about \speed{101.2}, and its maximum projected length and lifetime are about 8.5 Mm and 84 seconds, respectively. It is also measured that S10 is ejected at a speed of about \speed{93.9}. The time-distance plot also reveals the rotation motion of the filament threads during the ejection of the jets, which suggests the reconfiguration of the magnetic fields of the filament.

We use the Differential Emission Measure (DEM) technique to study the thermal dynamics and plasma density of S9. The DEM method utilizes optically thin multi-wavelength EUV images from the AIA instrument to calculate the emission measure and thereby infer the temperature and electron density distributions within the observed solar structures. We can use the following formula to calculate the weight-averaged temperature and density distribution of the selected region:

\begin{align}
T_{\text{EM}} &= \frac{\sum \left(\text{EM}_T \cdot T \right)}{\sum \text{EM}_T},\tag{1} \\
n &= \sqrt{\frac{EM}{L}},\tag{2}
\end{align}

where T represents the temperature, $\text{EM}_T$ denotes the emission measure at a specific temperature, and L is the depth or width of the structures along the LOS \citep{2018ApJ...856L..17S}. The weighted temperature map shows that the temperature of S9 is lower than its surrounding environment (see \nfig{fig4}(m)), suggesting S9 is a cooler feature relative to the high-temperature corona. We determined that the head of S9 has a weighted temperature ranging from logT=5.6 to 7.0, with an average value of 2.9 MK. The emission measure of the S9's head is measured to be \(2.6\times10^{28}\) cm\(^{-5}\) (see \nfig{fig4}(n)). Assuming S9 is cylindrical, with its length representing the height and its width as the diameter of the cylinder, the thickness in the LOS, assumed to be consistent with its width observed by the AIA, is approximately 7 arcsec or \(5.0\times10^8 \) cm. Combining this with the emission measure values, we calculated S9's electron number density to be \(7.2\times10^9\) cm\(^{-3}\). We then calculated the thermal energy density \(E_t=3n_ekT\) and the kinetic energy density \(E_k=\frac{1}{2}n_em_pv^2\) to estimate S9's total energy, where \(n_e\) is the electron number density, \(k\) is the Boltzmann constant, \(T\) is the temperature, \(m_p\) is the mass of the proton, and \(v\) is the velocity. The total thermal energy of S9 is estimated to be about \(1.358\times10^{27}\) erg, and the total kinetic energy is about \(9.644\times10^{25}\) erg. The total energy released by S9 is comparable to that of microflares and small coronal jets. Using the formula \(E_m=E_k+E_t=\frac{B^2}{8\pi}\) and assuming the thermal and kinetic energy were transformed from magnetic free energy through magnetic reconnection, we can calculate the magnetic field strength involved in the reconnection. The sum of the thermal and kinetic energy densities of S9 is 9.26 erg cm\(^{-3}\), which corresponds to a magnetic field strength of 15.25 Gauss. From the two misaligned filament threads shown in \nfig{fig4}(h), we can determine the angle between the two threads to be 11.02\degr. Assuming that S9 was driven by the component magnetic reconnection between the two crossing filament threads, the estimated magnetic field strength of 15.25 Gauss should be the transverse component of the magnetic field represented by the filament threads. Therefore, we can obtain that the magnetic field strength of the filament threads is about 80 Gauss based on the measured included angle. 

We list all the measured and derived parameters of all studied small jets in Table \ref{tab:table1}, including the occurrence time, length, width, density, temperature, velocity, lifetime, thermal energy, kinetic energy, energy density, and the transverse magnetic field strength involved in the component magnetic reconnection. It can be seen that the length of these small jets has a wide range of 1.9 -- 16.1 Mm, while the width ranges from 1.6 to 5.0 Mm. The density and temperature do not show many variations, and they are in the ranges of \(7.0 \times 10^{9} - 9.6 \times 10^{9}\) cm\(^{-3}\) and 2.31 -- 4.02 MK, respectively. All the velocity, lifetime, thermal energy, kinetic energy, and energy density of the small jets show a large range of variations; they are in the ranges of \speed{15.8 -- 113.1}, 36 -- 192 s, \(0.55 \times 10^{26} - 13.58 \times 10^{26}\) erg, \(0.28 \times 10^{24} - 96.44 \times 10^{24}\) erg, and 6.93 -- 14.39 erg cm\(^{-3}\), respectively. The transverse magnetic field component strength producing these small jets falls within the range of 13 -- 19 Gauss if we assume these small jets were all driven by the component magnetic reconnection between crossing filament threads. Assuming all studied small jets share the same included angle as S9, the magnetic field strength of the filament threads is estimated to range from 70 to 100 Gauss.

\begin{figure*}[thbp]
\centering
\includegraphics[width=0.82\textwidth]{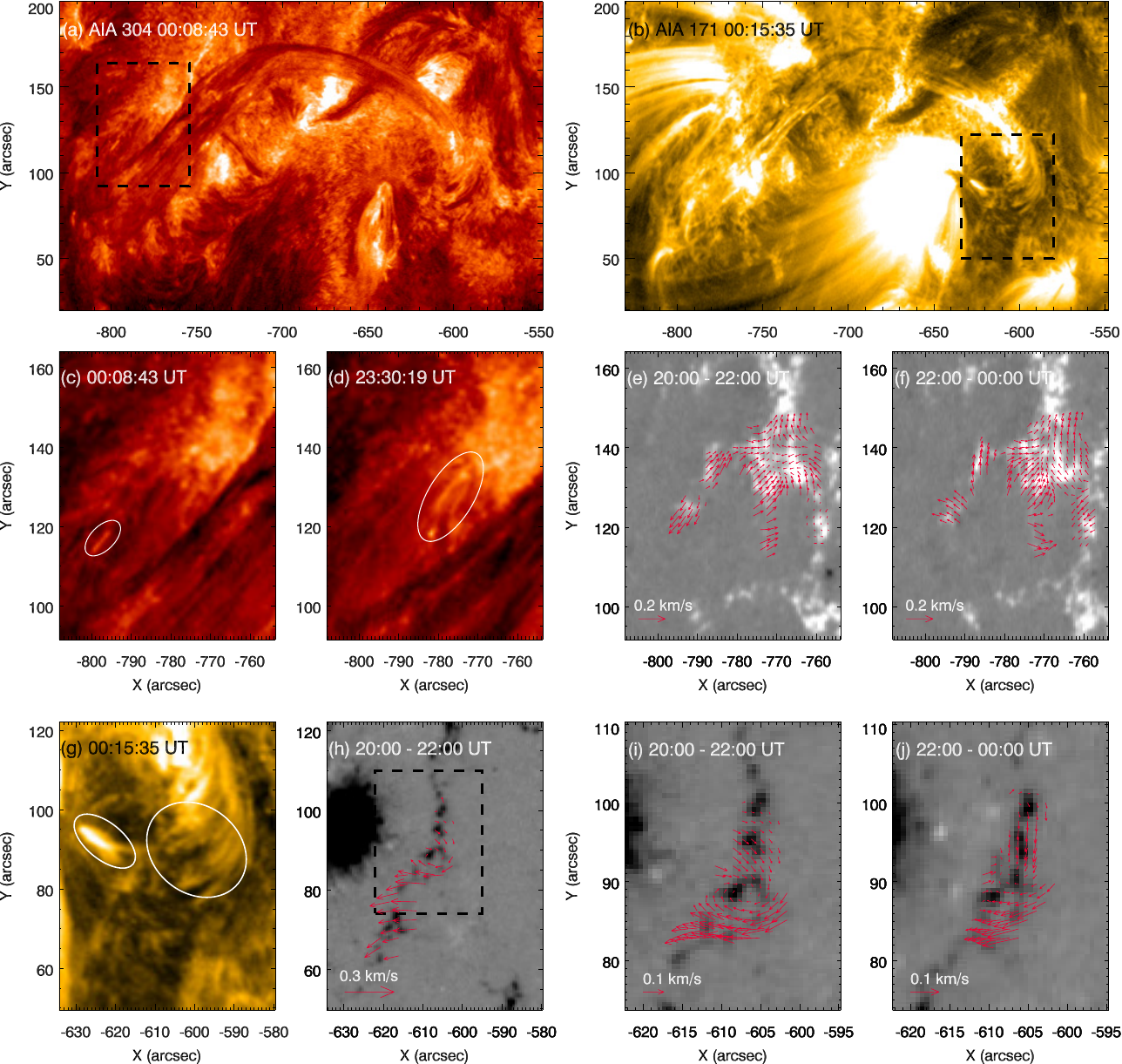}
\caption{
Results of the FLCT velocity field calculation. Panels (a) and (b) display an overview of the position for both footpoints. The black dashed boxes represent the fields of view for panels (c)-(f) and (g)-(h), respectively. Panels (c) and (d) show the location of the eastern footpoint, with circles indicating several brightening flows directed towards it. Combined with panels (e) and (f), these reveal that the eastern footpoint is situated within a strong positive-polarity network cell, and its corresponding velocity field is displayed in panels (e) and (f). Panel (g) shows the position of the western footpoint, where circles on the right highlight significant brightening flows converging towards it, while circles on the left indicate a portion of the flow entering the sunspot. Combined with panel (h), this suggests that the western footpoint roots in a negative-polarity network cell located west of the sunspot. The velocity field associated with the western footpoint is presented in panel (h), clearly exhibiting a vortex motion, the magnified views of which are shown in panels (i) and (j).}
\label{fig5}
\end{figure*}

To investigate what causes the loss of stability of the filament, we calculated the photospheric velocity map surrounding the east and west footpoints of the filament using HMI magnetograms with the Fourier local correlation tracking (FLCT) method \citep{2008ASPC..383..373F}. The eastern footpoint can be identified by tracking the end of the brightening spreads along the filament in AIA 304 \AA\ observations. As shown in \nfig{fig5} (a) (c) (d), both the filament structure and its associated brightening end near coordinates (-775, 135), corresponding to a strong positive-polarity network cell in the HMI magnetograms (see \nfig{fig5} (e) and (f)). Conversely, the western footpoint position is clearly indicated by the westward-propagating brightening flow (see \nfig{fig5} (b) and (g)). AIA 171 \AA\ observations demonstrate that the filament was anchored either in the western sunspot of the active region or at its western periphery, adjacent to the boundary of a negative-polarity network cell at approximate coordinates (-605, 90). We applied the FLCT method to derive the photospheric velocity fields near both footpoints using line-of-sight magnetograms from 29 August 20:00 UT to 30 August 00:00 UT, prior to filament splitting. The calculated velocity fields were averaged over two time intervals (20:00–22:00 UT and 22:00–00:00 UT), with analysis restricted to regions where the field strength exceeded 40 G at the eastern footpoint and was below -40 G at the western footpoint. Our results reveal that the positive-polarity fields at the eastern footpoint exhibit convergence motion toward the central strong magnetic network and demonstrate overall northward drift (see \nfig{fig5} (e) and (f)). In contrast, a localized portion of the negative-polarity fields at the western footpoint displays a distinct vortex pattern (see \nfig{fig5} (i) and (j)). These observations clearly show a clockwise-rotating vortex in part of the western footpoint, while no significant vortical motion is detected at the east footpoint. We therefore conclude that this filament exhibited footpoint rotation prior to eruption.

\begin{figure*}[thbp]
\centering
\includegraphics[width=0.85\textwidth]{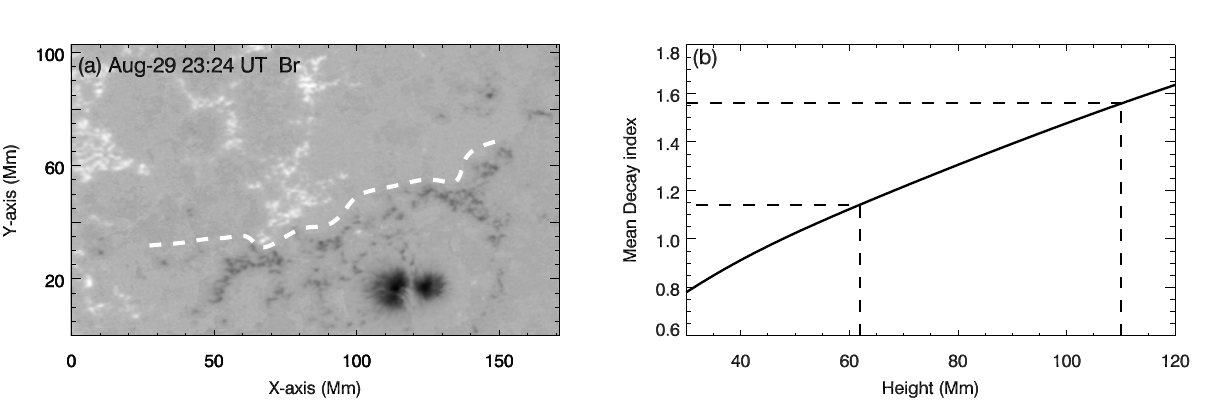}
\caption{
The dashed line in panel (a) indicates the orientation of the PIL, along which the decay index was calculated. Panel (b) presents the decay index values, averaged along the section of the PIL shown in panel (a), derived from the extrapolated magnetic field at 23:24 UT on August 29.}
\label{fig6}
\end{figure*}

We noted that the eruption of the upper filament branch did not show a clear rotation motion, suggesting that the eruption of the upper filament branch was unlikely triggered by the ideal kink instability \citep{2005ApJ...630L..97T}. Using the potential field source surface (PFSS) extrapolation method based on the code described in \citep{2012LRSP....9....5W}, we calculated the decay index of the horizontal magnetic field component at various heights above the PIL. The decay index, defined as \( n = -\mathrm{d} \ln B_h/\mathrm{d} \ln h \), where h is the height above the photosphere and \(B_h\) is the horizontal magnetic component, indicates the rate of magnetic field decay with height. When the decay index exceeds a certain critical value, it can trigger torus instability and lead to further expansion and potential eruption of the filament \citep{2006PhRvL..96y5002K}. Using a pair of 304 \AA\ images from {\em STEREO}/EUVI-B at 00:26:46 UT and {\em SDO}/AIA at 00:26:43 UT, we reconstructed the true height of the filament by using the procedure "\texttt{scc\_measure.pro}" developed by W. Thompson \citep{2012SoPh..276..241T}. The three-dimensional reconstruction result indicates that the upper branch of the filament reached a height of about 62 Mm at 00:26:46 UT; this height corresponds to a decay index value of about 1.14 that falls in the range of critical decay index values of 1.1–1.3 suggested in \citep{2010ApJ...718.1388D}, but is lower than the theoretical critical decay index value of 1.5-2.0 \citep{2006PhRvL..96y5002K}. This suggests that the upper branch of the double-decker filament at such a height could still be torus stable. We noted that in \nfig{fig2}(b), the projection height of the upper branch of the double-decker filament was at least 110 Mm above the limb of the solar disk at 01:40 UT on Aug 30, which marks the accelerated rise of cavity in \nfig{fig2} (c). In addition, the footpoints of the filament were rooted slightly on the backside of the Sun instead of on the disk limb. Therefore, the true height of the upper branch of the double-decker filament should be larger than 110 Mm. One can read from \nfig{fig6} that the decay index at the height of 110 Mm is larger than 1.55. Thus, the upper branch of the filament after the splitting was located at a height corresponding to the torus-unstable region. 

\section{Conclusions and Discussions}\label{sec:discussion} 
Based on stereoscopic high-resolution observations from {\em SDO} and {\em STEREO}, we discover that a double-decker filament can be formed when a braided MFR splits via internal component magnetic reconnection between the intertwined magnetic field lines. It's worth noting that during the splitting process, several phenomena occurred: the slow ascent of the filament, perpendicular small jets originating from brightenings on the filament threads, bidirectional mass flows starting from the brightenings and moving along the filament threads, and the rotational motion of the brightened filament threads. These characteristic phenomena serve as strong evidence supporting the occurrence of component magnetic reconnection between the intertwined magnetic field lines that make up the filament. The splitting of the filament lasted for approximately 107 minutes (23:28 -- 01:15 UT) and can be divided into two distinct stages - the weak splitting stage and the intense splitting stage. While the weak splitting stage aligned with both the intermittent brightening phase and the filament's slow inflation phase, the intense splitting stage coincided with the violent brightening phase and the rise of the upper branch. Following the splitting event, the upper branch of the double-decker filament erupted from approximately 01:30 UT. This started with a slow-rise phase, which was well fitted by a quadratic function but exhibited only minor acceleration. This phase persisted until the filament's disappearance at 01:55 UT, coinciding with the onset of the X-class flare. Correspondingly, at 01:40 UT, the overlying cavity entered its main acceleration phase, which was best described by an exponential function. These kinematic characteristics are consistent with the eruptive filament properties documented in active regions by \citep{2020ApJ...894...85C}. Subsequently, the upper branch erupted successfully due to the torus instability of the background confining magnetic field. This eruption was accompanied by a {\em GOES} soft X-ray M1.2 flare and a fast halo CME with a typical three-component structure.

The initiation of the component magnetic reconnection was probably driven by the slow rotational motion of part of the western filament footprint, which increased the magnetic twist of the filament and caused the differently driven parts of the flux to move and inflate in different ways, and, therefore, supports a splitting accompanied by reconnection. This reconnection mechanism is similar to the 3D MHD simulations by \cite{2025arXiv250201796C}, who demonstrated that rotational footpoint motions can induce magnetic twist in coronal flux tubes, leading to MHD avalanches characterized by widespread reconnections. \citep{2014ApJ...789...93C} observed a high-arching magnetic flux rope (MFR) anchored in/near a rotating sunspot that exhibited rise and inflation, demonstrating how single-footpoint rotation can drive filament motion. With the formation of the braided structure, magnetic reconnection occurred slowly and sporadically at first, probably only involving a small fraction of intertwined filament threads. This corresponds to the intermittent brightening phase as observed from 23:28:43 UT on August 29 to 00:08:31 UT on August 30. As the continuous accumulation of magnetic twists reaches a critical point, magnetic reconnection between some nearby magnetic field lines further leads to the avalanche of reconnections between other ones, resembling the occurrence of flares. This process corresponds to the violent brightening phase from 00:08:31 UT to 01:10:55 UT on August 30. Our analysis suggests that such a magnetic reconnection process not only released the accumulated magnetic twists in the filament but also led to the staged splitting and rising motion of the filament.

Drawing upon previous studies, it has been established that the splitting of a single filament is caused by magnetic reconnection that takes place within or above the filament, specifically between the two legs of the confining magnetic field lines \citep{2001ApJ...549.1221G}. Consequently, the onset of accompanying flares was anticipated to coincide with the start of the splitting process. For example, some recent studies indicated that the time interval between the splitting of the filament and the accompanying flare is about several minutes \citep[e.g.,][]{2012ApJ...750...12S, 2018ApJ...856...48C, 2023ApJ...953..148S}. Nevertheless, our observational analysis reveals that the splitting of the filament began more than two hours before the onset of the {\em GOES} M1.2 flare and the partial filament eruption. This suggests that the splitting mechanism in our case should be different from those proposed in previous studies. During the splitting process, the filament experienced a slow rising phase and was accompanied by numerous perpendicular small jets originating from the local brightenings on the filament threads, bidirectional mass flows starting from the brightenings and moving along the filament threads, and rotational motion of the brightened filament threads. All these phenomena indicate that magnetic reconnection was occurring within the filament. In addition, the direct observation of some misaligned filament threads in the AIA 171 \AA\ images also indicates that the filament should be a braided MFR consisting of different clusters of intertwined magnetic field lines, which suggests that such a braided MFR is liable to the occurrence of component magnetic reconnection at the crossings of the intertwined magnetic field lines. This is analogous to the situation in small-scale braided coronal loops, where magnetic reconnection is also prone to occur at the locations of magnetic crossings. In such a giant braided filament MFR, similar reconnection processes can take place. We propose that the splitting process in our study is independent of the subsequent eruption of the filament and the flare. After the splitting, the newly formed double-decker filament can remain stable or erupt. This is mainly determined by the magnetic environment in which the filament resides. Taking our event as an example, due to the torus instability of the background magnetic field, the upper branch finally erupted successfully and caused an M1.2 flare and a CME. However, if the background coronal magnetic field is torus stable and without other disturbances, the newly formed double-decker filament might remain stable for a long time. 

\cite{2017ApJ...841L..13C} reported many mini-jets almost vertically ejected relative to the axis of solar tornadoes caused by the interaction between filaments and EUV jets, which were accompanied by the untangling or unwinding of the highly twisted tornado structures. The authors elaborated on their study in a subsequent work \citep{2020ApJ...899...19C}; they further suggested that tornado mini-jets might be produced by two candidate reconnection processes, i.e., between erupting filaments and the surrounding magnetic fields or between the internal twisted or braided flux tubes in prominences. Different from those mini-jets observed in solar tornadoes, the small jets observed in the present event are obviously within the filament and were accompanied by the splitting of the filament. Our interpretation of the generation of these small jets is consistent with the one mechanism proposed by \cite{2020ApJ...899...19C}, i.e., due to the internal reconnection between braided flux tubes in prominences. The component magnetic reconnection at the crossing X-points, built by different clusters of magnetic field lines, cancels out the transverse component of magnetic fields and leads to the release of magnetic free energy and the transformation of the originally single braided magnetic flux rope into two separated ones with more parallel magnetic fields. A similar process, though on a smaller scale, has also been observed in recent studies of nanojets within braided coronal loops \citep{2021NatAs...5...54A, 2022ApJ...938..122P, 2022ApJ...934..190S}. 

Compared to other studies, small jets in the present study show similar characteristics, such as their size, density, and energy scales, which are typical of small-scale solar jets. Their lengths (1.9 -- 16.1 Mm) and widths (1.6 -- 5.0 Mm) are comparable to the dimensions of nanojets and mini-jets reported in other studies, such as those observed in braided coronal loops \citep{2021NatAs...5...54A, 2022ApJ...938..122P}. The densities ($7.0 \times 10^9$ -- $9.6 \times 10^9$ cm$^{-3}$) and temperatures (2.31 -- 4.02 MK) are similar to cooler features within the corona, as noted in previous works \citep{2022A&A...667A.166C}. Velocities (\speed{15.8 -- 113.1}) and lifetimes (36 -- 192 s) also match the ranges observed for small-scale jets \citep{2017ApJ...841L..13C}. Thermal energies ($0.55 \times 10^{26}$ -- $13.58 \times 10^{26}$ erg) and kinetic energies ($0.28 \times 10^{24}$ -- $96.44 \times 10^{24}$ erg) are consistent with microflares and small coronal jets \citep{2012ApJ...759....1G}. The energy density (6.93 -- 14.39 erg cm$^{-3}$) and transverse magnetic field components (13 -- 19 G) involved in reconnection are consistent with previous findings. These similarities highlight the role of component magnetic reconnection in driving small jets and filament splitting, contributing significantly to the energy dynamics of the solar atmosphere and reinforcing the importance of this mechanism in solar physics.

Taking advantage of the high spatiotemporal resolution observations available in recent years, component magnetic reconnection between braided magnetic field lines with small angles and the subsequent generation of small jets have become a hot research topic in current solar physics. It is linked to various solar activities such as localized heating in filament and coronal loops \citep{ 2022A&A...667A.166C,2023ApJ...958..116W, 2023A&A...679A...9B}, mini-jets in solar tornadoes \citep{2017ApJ...841L..13C, 2020ApJ...899...19C} and in braided coronal loops \citep{2021NatAs...5...54A, 2022ApJ...938..122P, 2022ApJ...934..190S,2025ApJ...982L..50W}, and it has been considered a potential candidate mechanism for addressing the coronal heating problem. The findings of the present study, combined with previous research, suggest that component magnetic reconnection is a ubiquitous process in the solar atmosphere, playing a crucial role in various solar activities, including filament splitting and the formation of double-decker filaments. The detailed analysis of the small jets' properties and their comparison with other studies reinforce the understanding of the underlying mechanisms driving these phenomena and their impact on the solar atmosphere's energy balance.

\begin{acknowledgments}
The authors thank the excellent data provided by the {\em SDO} and the {\em STEREO} teams. This work is supported by the Natural Science Foundation of China (12173083, 12273106), the ``Light of West China'' Program of the Chinese Academy of Sciences, and the Shenzhen Key Laboratory Launching Project (No. ZDSYS20210702140800001) and the Specialized Research Fund for State Key Laboratory of Solar Activity and Space Weather. The authors also gratefully acknowledge the constructive comments from the anonymous reviewers and the editor for their valuable contributions to the improvement of this manuscript.
\end{acknowledgments}

\end{document}